# How Tertiary Studies perform Quality Assessment of Secondary Studies in Software Engineering

Dolors Costal[0000-0002-7340-0414], Carles Farré[0000-0001-5814-3782], Xavier Franch[0000-0001-9733-8830], and Carme Quer[0000-0002-9000-6371]

Universitat Politècnica de Catalunya, Spain
{dolors, farre, franch, cquer}@essi.upc.edu

**Abstract. Context**: Tertiary studies are becoming increasingly popular in software engineering as an instrument to synthesise evidence on a research topic in a systematic way. In order to understand and contextualize their findings, it is important to assess the quality of the selected secondary studies. **Objective**: This paper aims to provide a state of the art on the assessment of secondary studies' quality as conducted in tertiary studies in the area of software engineering, reporting the frameworks used as instruments, the facets examined in these frameworks, and the purposes of the quality assessment. **Method**: We designed this study as a systematic mapping responding to four research questions derived from the objective above. We applied a rigorous search protocol over the Scopus digital library, resulting in 47 papers after application of inclusion and exclusion criteria. The extracted data was synthesised using content analysis. **Results**: A majority of tertiary studies perform quality assessment. It is not often used for excluding studies, but to support some kind of investigation. The DARE quality assessment framework is the most frequently used, with customizations in some cases to cover missing facets. We outline the first steps towards building a new framework to address the shortcomings identified. **Conclusion**: This paper is a step forward establishing a foundation for researchers in two different ways. As authors of tertiary studies, understanding the different possibilities in which they can perform quality assessment of secondary studies. As readers, having an instrument to understand the methodological rigor upon which tertiary studies may claim their findings.

## 1   Introduction

A tertiary study (TS) is defined as "a systematic review of systematic reviews" [1]. TSs are becoming increasingly popular in the software engineering (SE) field, because they offer the possibility to integrate existing knowledge that has been previously synthesised in secondary studies.

As in any other type of systematic review, assessing the quality of the secondary studies selected in a TS is a recommended activity to fully embrace the subject of investigation. Quality assessment (QA) of systematic reviews has been subject of investigation in the past but not in the specific case of TS. Therefore, a number of questions arise, e.g., what are the QA frameworks adopted by authors of TSs?, how is the QA outcome used in the analysis part of TSs? This paper presents the results of a systematic mapping exploring these questions.

## 2    Background and Related Work

Existing approaches from disciplines such as medicine and sociology with a long tradition of evidence-based research have influenced systematic reviews in SE, where the evidence-based paradigm has been adopted in the last fifteen years [2]. In the medicine discipline, there exist the quality criteria (QCs) proposed by the York University Centre for Reviews and Dissemination (CRD) Database of Abstracts of Reviews of Effects (DARE), which focuses on systematic reviews on health care interventions and services. A first version of the DARE QCs included four questions (DARE-4) [3] and, in 2008, they were extended to include an additional synthesis QC (DARE-5) [4]. DARE-4 and DARE-5 QCs are listed in Table 1; QC3 is only present in DARE-5. CRD uses those QCs to select reviews. In DARE-4, reviews have to meet at least three criteria, of which QC1 and QC2 are mandatory. In DARE-5, reviews have to meet at least four criteria of which QC1, QC2 and QC3 are mandatory.

**Table 1.** Quality criteria included in DARE

| QC | Description |
|---|---|
| QC1 | Were inclusion/exclusion criteria reported? |
| QC2 | Was the search adequate? |
| QC3 | Were the included studies synthesised? (not in DARE-4) |
| QC4 | Was the quality of the included studies assessed? |
| QC5 | Are sufficient details about the individual studies presented? |

In the SE area, DARE-4 was initially adopted by Kitchenham et al. [1, 5, 6]. In 2011, Cruzes and Dyba [7] proposed to use DARE-5, arguing that it is critically important to evaluate whether a systematic review synthesizes primary studies.
In some of Kitchenham et al. works [5, 6, 8], the DARE QCs are refined. More concretely, a detailed rubric to evaluate each question achievement is defined (to Yes/Partly/No), a score is assigned to each degree of achievement (1/0.5/0), and scores can be aggregated. On the other hand, these studies did not use the QA for systematic literature review (SLR) selection because that would remove "SLRs relevant to practitioners and educators" and because "even if an SLR is of relatively low quality, it might still provide a useful starting point for academics [...] as long as all the relevant primary studies are fully cited" [6].

Budgen et al. [9] developed a TS to analyze the reporting of systematic reviews and extracted a set of practical lessons on how to achieve the DARE QCs. They pointed out that the DARE QCs do not cover all quality issues: "the DARE criteria address what should be reported rather than how it should be reported [and...] they do not attempt to cover all issues [...]. For example, while we might expect any secondary study to include an assessment of threats to validity, this is not actually something identified as being a part of the DARE criteria."

In addition, some authors argue that there exist proposals that present advantages over DARE for assessing the quality of systematic reviews [10]. More specifically, they advocate the use of AMSTAR having its first version 11 quality items [11] and its second one 16 [12].

## 3    Research Method

This study takes the form of a systematic mapping following the guidelines defined by Kitchenham and Charters [1]. As a first step, we confirmed the need for the review by checking that there are not similar studies with the same aim as our paper. We only found a literature study by Yang et al. [13] that investigates QA performed in secondary studies, i.e., assessing the quality of primary studies. This is a significant difference with our focus on secondary studies, because primary studies and secondary studies are fundamentally different: while primary studies are really diverse, secondary studies should all fulfil some methodological criteria that have become well-established in the SE community. Therefore, the quality of primary studies and secondary studies cannot be assessed in the same way (although some similarities may still exist, see Section 5).

### 3.1    Research Questions

We have used the GQM methodology [14] to derive our research questions. Based on the GQM goal template, the goal of this study is to analyze the state of the art in TSs to characterize how they assess the quality of secondary studies from the point of view of empirical researchers in the context of SE research literature. The research questions are:

- RQ1. What are the demographics of the selected TSs?
- RQ2. What are the frameworks and their variations applied to assess the quality of secondary studies in TSs?
- RQ3. What facets are examined by the criteria applied to assess the quality of secondary studies in TSs?
- RQ4. What are the purposes of assessing the quality of secondary studies in TSs?

### 3.2    Search Protocol

In this research paper, we took the following decisions that impact on the search protocol:

i. Perform automatic search.
ii. Use the Scopus digital library as the only resource. As stated by Krüger et al., Scopus indexes papers of main publishers as ACM, IEEE, Springer and Elsevier [15].
iii. Restrict to papers written in English and published since 2004 in the SE area. Year 2004 was the publication date of the seminal paper on Evidence-Based Software Engineering [16].

We applied the following steps:

1. Design and pilot the search string in Scopus. We ended up with "TERTIARY STUDY" as search string. To implement it in Scopus, we also fine-tuned the Scopus search parameters. For instance, we realized the importance not to restrict the Document Type, given that some TS published in journals are classified as Report instead of Article. Also, we decided to select "Computer Science" and "Engineering" as subject areas since SE papers could eventually fall in any of them. The final Scopus search string can be found in our replication package[1].
2. We executed the final search string over title, keywords and abstract with date 19-January-2021, resulting in 103 candidate papers.
3. We applied the inclusion criteria (IC) and exclusion criteria (EC) presented in Table 2 over title, keywords and abstract of these 103 papers, excluding 46 of them. IC and EC were applied by two team members to every paper, with the agreement not to exclude papers in case of doubt. Seven out of the 103 papers were conflictive, but after a plenary meeting, the team arrived to full consensus.
4. Later on, when analysing the 57 remaining papers, we excluded 10 additional works when the full text showed that they did not fulfil some IC or violated some EC. Exclusion was proposed by the team member in charge of extracting the data, and discussed and agreed upon in a plenary meeting. This led to 47 TSs selected to analyse.

**Table 2.** Inclusion criteria (IC) & exclusion criteria (EC)

| | |
|---|---|
| IC1 | The paper is a tertiary study providing a survey of secondary studies in the SE field |
| IC2 | The paper is published from 2004 onwards |
| IC3 | The paper is written in English |
| EC1 | The paper is superseded by a later version from the same team of authors |
| EC2 | The paper describes the TS with very little detail, making its analysis unfeasible |
| EC3 | The paper is not available (even after contacting authors) |

### 3.3 Data Extraction, Analysis and Reporting

We stored the result of the search in a GDrive spreadsheet, which was used in the rest of the study (shared as a replication package). Selected TSs are numbered [S01]-[S47]. We kept track of excluded papers for traceability reasons. We added as many columns as needed to extract the data required to answer the RQs. One team member was in charge of extracting the data for every individual paper; the set of papers were split into the team members at equal share. We held weekly plenary meetings to analyse progress and discuss issues as they emerged.

We used content analysis to synthesise codes from the extracted data. Inductive coding prevailed, although we combined it with deductive coding in RQ3. In general, we paid attention to avoid researcher bias through several actions: working in pairs, supervision (i.e., a researcher validating results from another), weekly plenary meetings and explicit check of inter-rater agreement when we thought it was needed.

---

[1] Replication package at Zenodo: https://doi.org/10.5281/zenodo.4742147

## 4 Results

### 4.1 RQ1. Demographics

**Prevalence of QA.** Twelve out of the 47 selected TSs (25.5%) do not assess the quality of the secondary studies they included. Four of them claim that the applied search/selection procedure guarantees the good quality of the included secondary studies by either: limiting the set of venues where they are published [S16, S32], or obtaining them through an index like Scopus [S21], or requiring them to follow Kitchenham's guidelines [1] [S20]. The remaining 8 TSs do not mention this issue.

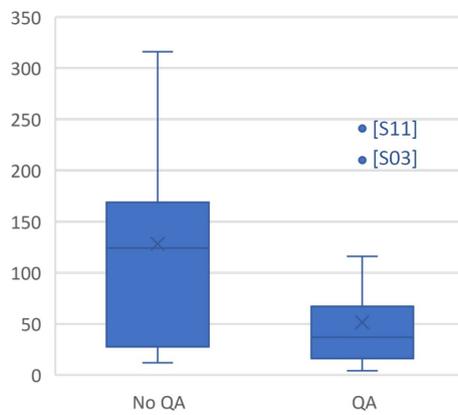

**Fig. 1.** Number of included secondary reviews in TS.

**TS Types.** From the thematic analysis of the TSs, we observed that there are two main types as follows: (i) *SE-Area*: TSs that investigate the state of the research in a specific SE area; (ii) *Methodological*: TSs that focus on the methods and protocols followed by secondary studies in their development process.

Accordingly, 25 out of the 47 selected TS (53.2%) are *SE-Area* and 22 (46.8%) are *Methodological*. Among the different SE areas addressed, Agile Software Development is the most recurring one, with 4 TSs, followed by Global Software Development with 3 TSs, and Software Product Lines and Software Testing, with 2 each. In the case of [S03], its SE area encompasses the whole SE field. Among the Methodological studies, four specific topics are subject of analysis by more than one TS: QA of Primary Studies, Types of secondary studies to consider, Search, Threats to Validity, and Synthesis. We labelled 5 *Methodological* TSs as *Generic*, in the sense that they cover several methodological topics at once.

**Types of secondary studies.** Seven TS papers considered only systematic literature reviews (SLRs), 1 paper [S03] only systematic mappings (SMs), 22 TSs papers both SLRs and SMs, 2 TSs only SLRS and Meta-analysis, and the remaining 15 considered more types (SLRs, SMs, Meta-analyses, Surveys, etc.).

**Number of secondary studies.** The average number of secondary studies per TS is 71,1. However, we observe disparities in the number of secondary studies considered when partitioning the 47 TSs into two groups: those that do not perform QA and those that do perform QA. The first group ("No QA" in Fig. 1) includes 12 TSs that are all *Methodological* but one. The second group ("QA" in Fig. 1) includes 35 TSs, 11 of which are *Methodological*, and the remaining 24 *SE-Area*. The two resulting box-plot diagrams shown in Fig. 1, where two outliers are made explicit, illustrate the clear difference between the two groups.

### 4.2  RQ2. Frameworks used in QA

**QA framework.** Most of the 35 papers reporting QA adopt an existing framework either directly or customizing it by adding or removing QCs. There are a few, i.e., 4 papers, that build their own custom-made QCs. In the former case, as we can see in Fig. 2, the framework chosen is DARE-4 or DARE-5 except for one study that customizes the AMSTAR framework. We can see that 25 out of the 31 studies adopting an existing framework apply it directly while 6 perform some adaptations. All adaptations consist of adding QCs to DARE-4 or DARE-5 except for one study that customizes AMSTAR by removing one criterion.

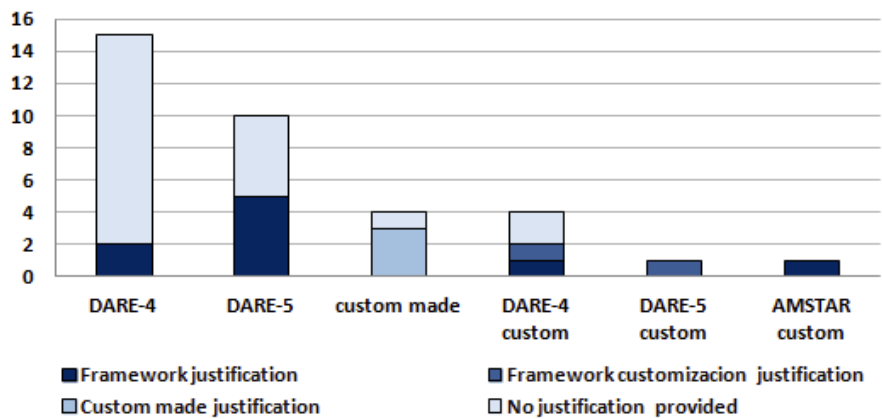

**Fig. 2.** Quality assessment frameworks

**QA framework historical evolution.** In the first years of the search period until 2015 all papers reporting QA used DARE-4 or DARE-5 directly with the single exception of a study from 2013 that defines custom-made criteria (1 out of 8). There is a turning point from 2015 onwards because in this latter period one third of the papers (9 out of 27) use QCs not present in DARE. Five studies customize DARE-4 or DARE-5 by adding QCs to them, three studies build custom made QCs and one adopts the AMSTAR framework which is more extensive than DARE.

**Justification of the selection.** Only 40% of the studies reporting QA (14 out of 35) provide a justification either for: (1) their framework choice, or (2) their framework customization, or (3) their custom-made approach. Each of them gives a single justification for only one of those aspects. The stacked bars in Fig. 2 depict the number of papers giving each kind of justification for each type of QA. We can see that the largest proportion of papers not providing justifications are those adopting the DARE-4 framework directly, i.e., 13 out of 15.

There are 9 papers that justify their framework choice and all of them argue about the good acceptance of the framework, except for [S24] that selects AMSTAR for its reliability since QA "requires the use of a valid and reliable tool" and for [S36] which also provides reliability arguments for their choice of DARE-5. There are 2 papers giving a justification for customizing a framework: [S14] mentions the need of adding missing aspects to DARE-4 and [S15] mentions the need of adapting DARE-5 to their specific study. Finally, 3 out of the 4 papers building their custom-made QCs provide explanations: 2 papers mention the need to adequate the QCs to their study research and paper [S08] says they have established their QCs to cover the four main areas of empirical research according to [17].

**QA Measurement.** Regardless of the QA framework applied, the 35 TSs with QA can be classified into three main types according to how they measure quality:

- *Aggregator*: 27 TSs aggregate the scores obtained from appraising each QC. Remarkably, all the QCs always have the same weight. All these 27 TSs but one summed the individual scores; [S10] used the arithmetic mean instead.
- *Non-Aggregator*: Five out of the 35 TSs that perform QA appraise each QC but do not aggregate the resulting scores.
- *Unknown*: Three TSs, [S9, S22, S29], do not provide any information on how quality is assessed beyond stating that they use a DARE framework.

**QC Appraisal.** The 32 TSs (27 *Aggregator* + 5 *Non-Aggregator*) have different ways of appraising each QC:

- *Uniform Three-Value*: 24 TSs measure each QC using three possible values, usually "Yes", "Partially" and "No", which are then mapped into 1, 0.5, and 0, if they are aggregated.
- *Uniform Other Values*: Three TSs use two values, "Yes" and "No", for each QC; and [S24] assesses its list of customized AMSTAR criteria with five possible values: "Yes", "No", "Partially", "Cannot answer", "N/A".
- *Non-Uniform*: 4 TSs use a three-value score (e.g. 0, 0.5, 1) for some QC, and for the rest, they use a binary one (0, 1).

**Breakdown of QC scores**. The 32 TSs (27 *Aggregator* + 5 *Non-Aggregator*) can also be classified into three types:

- *Breakdown*: 23 TSs report the scores for each QC and secondary paper. Two of them include these breakdown scores in the supplementary material.
- *No Breakdown*: 9 TSs do not report any breakdown scores. Here we single out [S11, S38], which do not provide the breakdown for each secondary study but instead report the count of secondary studies for each possible score/QC.

### 4.3 RQ3. Facets examined in QA

**Individual QCs**. The 35 TS that perform QA of secondary studies reported a total of 181 QC altogether. We induced 31 different codes from these 181 QC, see Table 3.

Not surprisingly, the most used codes corresponded to the four DARE-4 criteria (32-33 occurrences each) with the additional synthesis criteria from DARE-5 coming next (QC3, 11 occurrences). Still, it is worth mentioning that some papers provided the definition of these five criteria in their own words. The rest of the codes, proposed in 10 TSs, were spurious, with only QC6 exceeding 3 occurrences.

**Table 3.** Q3 Coding

| Code id. | Topic | #TSs |
|---|---|---|
| QC1 | Inclusion & exclusion criteria (DARE-4, -5) | 32 |
| QC2 | Search coverage (DARE-4, -5) | 32 |
| QC3 | Studies synthesized (DARE-5) | 11 |
| QC4 | Quality assessment (DARE-4, -5) | 33 |
| QC5 | Studies reporting (DARE-4, -5) | 33 |
| QC6 | Statement of goal and RQs | 4 |
| QC7 | Statement of RQs using PICOS | 1 |
| QC8 | Inclusion of validity threats | 2 |
| QC9 | Mitigation of bias | 2 |
| QC10 | Traceability of QA into primary studies | 1 |
| QC11 | Traceability of evidence into primary studies | 1 |
| QC12 | Usage of QA and evidence in synthesis | 2 |
| QC13 | Conclusions relying on evidence | 1 |
| QC14 | Aggregation of results in synthesis | 1 |
| QC15 | Weighting results during synthesis | 1 |
| QC16-20 | Five additional detailed criteria in QA | 1 |
| QC21 | Review supported by case study or survey | 1 |
| QC22 | Type of systematic review | 2 |
| QC23 | Adequacy of the research design | 3 |
| QC24 | Team involved in selection & data extraction | 1 |
| QC25 | List of included & excluded studies provided | 1 |
| QC26 | Declaration of any conflict of interest | 1 |
| QC27 | Focus on the scope of the study | 2 |
| QC28 | Precise reporting of the findings | 3 |
| QC29 | Precise reporting of research method | 3 |
| QC30 | Value of the work made evident | 1 |
| QC31 | Protocol design | 1 |

**Categorization of codes**. Next, we categorized these codes into facets. We consolidated the phases of systematic reviews proposed in Kitchenham et al.'s works [1, 2] into a single classification schema, removing some steps or concepts we considered not related to QA (e.g., commissioning the review). We made this consolidation through dedicated meetings. We obtained 11 categories grouped into the three usual stages: Planning, Conducting and Reporting. We aligned this classification schema with the 31 codes, not allowing multiple categorization (we chose the category that we considered dominant for each code). In this step, we added 5 new categories and decomposed the single Reporting category into 6 more fine-grained categories. We finally obtained the classification schema shown in Table 4, with 21 categories, from which three of them did not have any code associated. At the end, a majority of 125 out of the 181 QA criteria reported correspond to the Conducting stage (69.1%), while only 12 correspond to Planning (6.6%) and the remaining 44 (24.3%) to Reporting.

**Table 4.** Categorization of QC codes

| Id | Category | QA codes |
| --- | --- | --- |
| P1 | Identify the need for a review | -- |
| P2 | *Determine scope* | QC27 |
| P3 | Specify goal and RQs | QC6, QC7 |
| P4 | *Design research method* | QC21, QC23 |
| P5 | Develop protocol | QC31 |
| P6 | Validate protocol | -- |
| C1 | Search studies | QC2 |
| C2 | Select studies | QC1 |
| C3 | Assess quality | QC4, QC16-QC20, QC22 |
| C4 | Extract data | -- |
| C5 | Synthesise | QC3, QC12, QC14, QC15 |
| C6 | *Mitigate bias* | QC9 |
| C7 | *Involve the team adequately* | QC24 |
| C8 | *Extract conclusions* | QC13 |
| C9 | Analyse limitations | QC8 |
| R1 | *Statement of CoIs* | QC26 |
| R2 | *Reporting of research method* | QC29 |
| R3 | *Reporting of findings* | QC28 |
| R4 | *Description of studies* | QC5, QC25 |
| R5 | *Traceability into studies* | QC10, QC11 |
| R6 | *Evidence of value* | QC30 |

### 4.4 RQ4. Purposes of QA

In this research question, we used deductive coding. The coding scheme is an extension of QA purposes in [1] proposed by Yang et al. [13] to analyse SE secondary studies QA purposes.

Most of TSs (23 out of 35) use QA for only one purpose, while six and four TSs use QA for two and three purposes respectively. Note that two TSs do not declare any purpose for doing QA.

Fig. 3 shows the percentages of TSs that use QA for each purpose (dark blue bars). More than half of the 35 TSs (60%) used QA to understand the quality of the secondary studies (*Investigation*), normally through a research question. Eight studies used QA results as additional IC/EC (*Selection*). The rest of purposes are gradually decreasing in application, and remarkably the three least declared purposes seem to be the most complex to carry out: use of QA for determining the strength of the inferences (*Interpretation*), for weighting the studies during synthesis (*Weighting*), and for explaining study results according to quality differences (*Differentiation*).

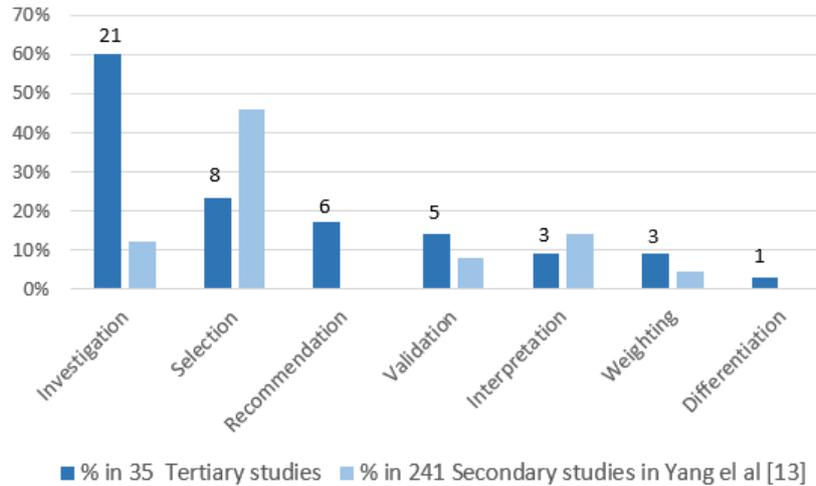

**Fig. 3.** Percentages of studies per purpose

## 5 Discussion

**QA in TSs versus QA in secondary studies**. Although QA in TSs and secondary studies is very different in nature, still we can compare the purposes of each, which have been analysed by Yang et al. [13], see Fig. 3. The differences show that the main purpose of QA in secondary studies is *Selection* of primary studies, far more than any other purpose, as shown in Fig. 3. This means that secondary studies use QA to ensure good quality of the primary studies analysed to synthesise the state of the art of a specific domain. In contrast, *Selection* of secondary studies is a purpose only in 23% of TSs, showing that authors of TS seem to consider that secondary studies are relevant for a TS independently of their quality. This idea is corroborated in [S01, S04, S15, 6]. Conversely, secondary studies do not investigate the quality of primary studies to support any research goal as often as in the case of TSs; we argue that this is due to the intrinsic diversity of primary studies.

**Impact of the type of TS in QA**. We have investigated further the influence of the TS type into our findings. Some results seem to point out that methodological TSs are less strict in assessing the quality of secondary studies than SE-area TSs. Almost all the TSs not performing QA are methodological (11 out of 12 TSs), and this is also the case of the two papers that do not state the QA purpose. A possible explanation is that methodological TSs focus on analysing in depth a particular aspect of the literature review process of the secondary studies included in the study and thus do not pay much attention to their quality, specially when there is a limited number of pages to report the study as it happens in conference papers, which is the case of 10 of the 22 mentioned methodological papers. Anecdotally, we remark that our own paper provides evidence related to this hypothesis! Another interesting observation is the greater number of secondary studies included in methodological TSs, which can be justified because they do not restrict to a particular area, therefore all secondary studies are likely to be candidates for inclusion. A last observation is that only 4 out of the 10 TS including non-DARE-related codes are methodological.

**Framework**. Nearly three quarters of the studies reporting QA (71,4%) adopt DARE-4 or DARE-5 frameworks directly. All remaining studies (10 TSs) use QCs not present in DARE, mostly by adding QCs to DARE-4 or DARE-5 (in 5 TSs), but also by building custom made QCs (4 TSs), or by adopting the AMSTAR framework which is more extensive than DARE (1 TS). This trend is more pronounced from 2015 onwards as mentioned in Section 4.2. The justifications given by the studies that customize DARE are the need of covering missing aspects or adapting the framework to their specific study. Furthermore, as seen in Section 2, Budgen et al. point out that DARE does not cover all quality issues [9] and we can see in Table 4 that there are many phases of systematic reviews not covered by it. We refer to Section 6 for the first steps towards this direction.

**Unclear purpose of QA in TSs**. The answer to RQ4 shows a great diversity of purposes for which researchers assess the quality of secondary studies in TSs. The most popular purpose is *Investigation*, but in a deeper analysis of the TSs with this purpose, most of them used the result of QA as if it were yet another bibliometrics measure, reporting the numbers in more or less detail and adding a reflection on how the obtained score measures the quality of the work. Besides being a conservative approach, it suffers from the risk pointed out by Budgen et al. in reference to DARE: QA frameworks are only concerned with whether QA has been done, not how well it has been done [9]. For this reason, we argue that more insightful purposes such as *Interpretation*, *Weighting* or *Differentiation* should be adopted more often in TSs. In any case, selecting one purpose or another should be a logical consequence of some well-established goal.

## 6   Towards a Comprehensive QA Framework

RQ3's results and the discussion above point to the direction that it would be positive to have a comprehensive framework covering all phases of systematic reviews and then customizations would rather be to cut the criteria that should not be applied in a particular study reducing the need of authors to perform their own extension of a framework. In this section, we outline how this new framework proposal can be developed from the results of the present study.

This new framework, QUASY (QUality Assessment of SecondarY studies), is built in the following manner according to the categorization that we propose in Table 4:

- Given the wide acceptance of the DARE framework, categories C1-C3, C5 and R4 shall be based upon the DARE formulation. However, we acknowledge that some of the proposed works contribute to make these DARE's QC closer to the needs of TS in SE, therefore we need to analyse whether these proposals are worth to be incorporated into the DARE formulation. For instance, QC25 could be incorporated into QC5 by requiring the list of studies to be provided in order to achieve the maximum score in the evaluation of R4 (Description of Studies).

- Those categories not addressed by DARE, but addressed by some of the studies found, shall be based upon the consolidation of the QCs elicited in this study. This concerns categories P2-P5, C6-C9, R1-R3 and R5-R6. For instance, the way in which S24 states QC31 ("Has the preliminary design of the research (i.e. the definition of research questions and the inclusion criteria) been correctly stated before starting the research?") seems a good starting point for category P5 (Develop Protocol), although we think that the protocol should include additional elements and thus the final formulation needs dome work.

- Last, those categories for which we have not found any QC in the literature, deserve special care in order to come up with solid proposals. Categories P1, P6 and C4 are in this situation. We have elaborated this novel part of our proposal in more depth and provide concrete proposals below, including score definition based on the guidelines of Kitchenham et al. [1, 2]:
  - **P1: Need for a review**. *Is the justification of the need for a review described and appropriate?* Score: *Yes*, a) the need for a study on the topic is duly argued, particularly if there exist relevant related studies on the same topic, which must be cited properly. *Partly*, the need for a review is not clearly justified or relevant related studies are not cited. *No*, the need for a review is not justified and no relevant related study is cited.
  - **P6: Validate protocol**. *Is the validation of the review's protocol described and appropriate?* Score: *Yes*, the review described explicitly how it was confirmed, prior to the execution of the protocol, that a) the search strings were appropriately derived from the research questions, b) the data to be extracted would properly address the research question(s), and if the data analysis procedure was appropriate to answer the research questions. *Partly*, some aspect (a, b, or c) is missing or all three are present but not addressed appropriately. *No*, the validation of the protocol is not explicitly described and cannot be readily inferred.
  - **C4: Data extraction**. *Is the review's data extraction process described and appropriate?* Score: *Yes*, a) the data extraction form is described explicitly, as well as b) the strategy for extracting and recording the data, and c) how the data extraction process was undertaken and validated, particularly when data required numerical calculations or were subjective. *Partly*, some aspect (a, b, or c) is missing or all three are present but not described with sufficient detail. *No*, the data extraction process is not defined and cannot be readily inferred.

The QUASY framework should also define scores in a consistent manner, and include a well-defined score aggregation mechanism.

## 7 Threats to Validity

*The study does not include all published TSs on SE*. The cause of this threat is: 1) We used only one digital library; 2) We did not apply snowballing. To understand the severity of the first threat cause, we evaluated the coverage of Scopus. We conducted equivalent searches in Scopus, IEEE Xplore, ACM DL, SpringerLink, ScienceDirect and WoS on April 28th, 2021, obtaining 106, 22, 13, 24, 2 and 36 papers, respectively (details can be found in[2]). Once analyzed, only 5 papers were not found by the Scopus search: 2 of them not written in English, other 2 were published in journals not indexed in Scopus (SIGSOFT SEN and SIGCSE Bull.) and the last one was a paper indexed in Scopus whose area was not Computer Science nor Engineering. Therefore, none of these 5 papers were relevant for our study. Regarding the second threat cause, before facing the cost of snowballing, we wanted to have initial results on the subject of investigation to uncover any possible new research question of interest that could be the subject of future work.

*The study is not fully replicable*. This threat is inherent to any literature review using digital libraries, due to their ever-changing nature. To mitigate this threat, we have documented in depth all aspects of our search process, including search string, parameters and dates, and have made all data available in a replication package.

*Results may be biased by researchers' judgement*. We mitigated this threat by intensive teamwork, e.g.: 1) we had weekly 90-minute plenary meetings to discuss relevant aspects of the research methods, results and analysis; 2) we discussed, and resolved offline through GDocs comments, minor discrepancies; 3) we did not exclude any paper without consensus of all researchers; 4) we intensively discussed all coding aspects through an iterative approach, processing the surveyed TS by batches and reflecting on the coding after every batch.

## 8   Conclusions

In this paper, we have investigated the current state of the art on quality assessment (QA) of secondary studies as conducted in tertiary studies (TS). Our main conclusion is that, in spite of the great amount and increasing maturity of TSs in the SE discipline, there is a lack of a comprehensive framework covering all facets of quality, with a clear rationale to decide the most appropriate QA purpose according to the goal of the TS. As response to these shortcomings, we have used the results of this literature review as the basis to outline a comprehensive and rationalized first version of the new QUASY framework for secondary studies. Our future research aims at developing in full and validating this framework. Our long-term aim is to provide a consolidated framework similar to that proposed by Ampatzoglou et al. [18] for the assessment of threats to validity. The QUASY framework shall include also methodological aspects, for instance concrete advice on the QA purpose (cf. RQ4) in relation to the main objective of the TS in which the QA is conducted.

## References

References [S01]-[S47] to TSs are in https://doi.org/10.5281/zenodo.4742147 (replication package at Zenodo).

---

[2] http://www.upc.edu/gessi/rep/Scopus-Coverage-Validation.xlsx